\documentclass[12pt,letterpaper]{article}

\usepackage[margin=1.25in]{geometry}
\usepackage[english]{babel}
\usepackage[T1]{fontenc}
\usepackage[]{times}
\usepackage{amsmath,amssymb,amsthm,stmaryrd}
\usepackage{bbm}
\usepackage{natbib}
\usepackage{graphicx}
\usepackage{comment}
\usepackage{array}
\usepackage{longtable,dcolumn,booktabs,mathtools}
\usepackage{url}
\usepackage{pdflscape}
\usepackage{pdfpages}
\usepackage{setspace}

\usepackage{caption}
\usepackage{dcolumn}
\newcolumntype{d}[1]{D{.}{.}{#1}}

\newcommand{\dd}{\textrm{d}}
\newcommand{\one}{\ensuremath{\mathbbm{1}}}
\newcommand{\na}{\ensuremath{_{\textup{NA}}}}

\newcommand{\hsp}{{\hspace{0.5mm}}} 

\usepackage{natbib}
\begin{document}

\begin{center}
{

\LARGE \bf 

Predicting Inflation: Professional Experts Versus No-Change Forecasts

} 

\bigskip
\begin{singlespace}
{\bf \large Tilmann Gneiting, Thordis L.~Thorarinsdottir}

\smallskip
{\bf Institut f\"ur Angewandte Mathematik \\ Universit\"at Heidelberg, Germany}

\bigskip
{\bf \today}
\end{singlespace}
\end{center}

\vspace{1mm}
\begin{abstract}
We compare forecasts of United States inflation from the Survey of
Professional Forecasters (SPF) to predictions made by simple
statistical techniques.  In nowcasting, economic expertise is
persuasive.  When projecting beyond the current quarter, novel yet
simplistic probabilistic no-change forecasts are equally competitive.
We further interpret surveys as ensembles of forecasts, and show that
they can be used similarly to the ways in which ensemble prediction
systems have transformed weather forecasting.  Then we borrow another
idea from weather forecasting, in that we apply statistical techniques
to postprocess the SPF forecast, based on experience from the recent
past.  The foregoing conclusions remain unchanged after survey
postprocessing.

\end{abstract}

\section{Introduction}  \label{sec:introduction} 

A wealth of societal decisions can benefit from accurate forecasts of
future inflation, ranging from the setting of monetary and fiscal
policies to negotiations of wage contracts and investment judgments.
To predict inflation rates, various methods have been employed,
including statistical time series techniques, methods based on term
structures and yield curves, and survey-based measures from consumers
or professional experts.  A prominent recent study argues that survey
forecasts perform best \citep{Ang&2007}.

\begin{figure}[t]

\centerline{\includegraphics[width=\textwidth]{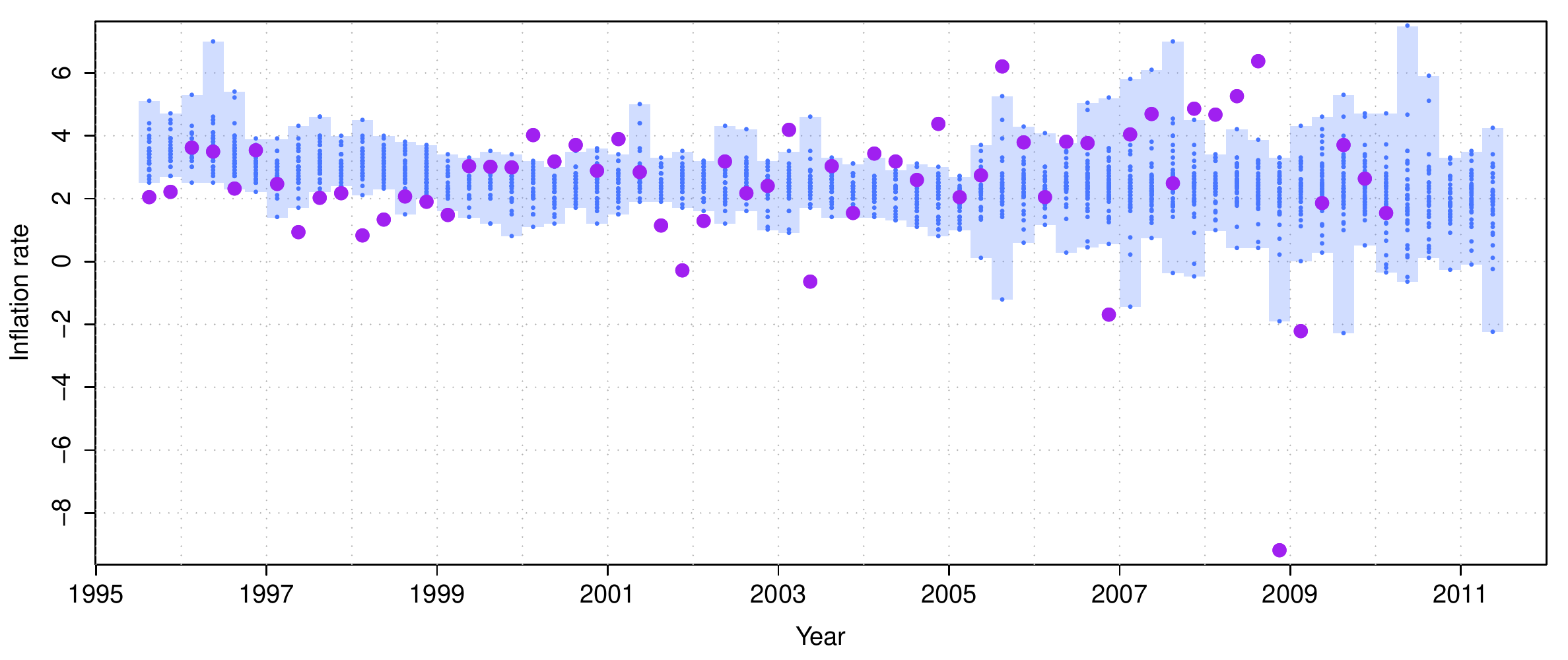}}

\vspace{-1mm}
\caption{SPF forecasts of United States inflation at a prediction
  horizon of five quarters.  The small blue dots show the individual
  SPF experts' point forecasts, with the pale blue boxes indicating their
  range.  The large purple dots represent the observed inflation
  rates.  Note the extreme deflation in the fourth quarter of
  2008. \label{fig:SPF} }

\end{figure}

The Survey of Professional Forecasters (SPF) is the leading quarterly
survey of macroeconomic variables in the United States (Zarnowitz,
1969; Croushore, 1993).  \nocite{Zarnowitz1969, Croushore1993} It
began in 1968 and was conducted by the American Statistical
Association and the National Bureau of Economic Research, before the
Federal Reserve Bank of Philadelphia took over in 1990.  The panel
comprises university professors as well as private sector economists,
who are asked each quarter to predict a range of macroeconomic
variables for the current and each of the following four quarters.
Our interest here is in inflation, in the form of the
quarter-over-quarter change of the consumer price index expressed in
annualized percentage points, which we now refer to as an inflation
rate.  Figure~\ref{fig:SPF} displays observed United States inflation
rates since the third quarter of 1995 along with the SPF forecasts
issued a year earlier.  Following common practice, we talk of a prediction
horizon of one quarter when referring to current quarter nowcasts, and
to prediction horizons of two to five quarters for the following four
quarters \citep{CroushoreStark2001}.  For example, forecasts issued a
year earlier correspond to a prediction horizon of five quarters.

The SPF panel forecast can be summarized to provide a single point
forecast, where we also follow common practice and take it to be the
median of the individual experts' predictions \citep{Stark2010}.  The
traditional no-change forecast equals the most recent available
observed rate.  This is the classical reference forecast in the
economic literature and that used by the Federal Reserve Bank of
Philadelphia.  Here, we introduce a novel kind of simplistic
reference forecast, which we call the probabilistic no-change
forecast.  It considers a rolling training period, consisting of the
20 most recent observed inflation rates, and takes the median thereof
as a point predictor.  This can be interpreted as a theoretically
optimal point forecast under a white noise assumption for the
inflation rates, whereas the traditional no-change forecast can be
interpreted as optimal under a random walk model.  In Section
\ref{sec:reference} we compare the predictive performance of the SPF
point forecast to the simple no-change forecasts.  Essentially, in
current quarter nowcasts economic expertise is persuasive.  At
prediction horizons beyond the current quarter, the SPF forecast
outperforms the traditional no-change forecast, but not the novel, 
equally simplistic probabilistic no-change forecast.

The SPF forecast can also be interpreted as an ensemble forecast,
similar to the ways in which weather and climate scientists have been
using ensemble prediction systems with great success
\citep{Palmer2002}.  State of the art weather forecasting uses
ensembles whose members are point forecasts from numerical weather
prediction models, with the members differing in initial conditions
and/or the specifics of the numerical model used.  For ensemble
weather forecasts, some form of statistical postprocessing is required
to correct for model biases and insufficient representations of the
forecast uncertainty \citep{GneitingRaftery2005}.  From
Figure~\ref{fig:SPF}, we see that similar to weather ensembles the SPF
ensemble forecast is uncalibrated, in that too many observations fall
outside the range of the ensemble forecast.  

Postprocessing methods in meteorology provide statistically 
corrected predictive probability distributions
for future weather quantities that condition on the ensemble
forecast.  From the predictive distribution, the probability of any
event of interest can be computed, and one can issue the optimal point
forecast under the loss function at hand \citep{Diebold&1998,
  Engelberg&2009}.  We adopt this approach and develop statistical
postprocessing methods for the SPF by using heteroscedastic regression
and Gaussian mixture models.  In Section \ref{sec:postprocessing}
these methods are introduced in detail and their predictive
performance is evaluated, with results that resemble those without
postprocessing.
Finally, in Section \ref{sec:robust} we study the robustness of our
results.  Concluding remarks are given in Section
\ref{sec:discussion}, where we discuss the methodological as well as
the economic and societal implications of our work.

\section{Predictive performance: SPF versus no-change forecasts}
\label{sec:reference} 

The Survey of Professional Forecasters (SPF) panel comprises academic
as well as private sector economists, who are asked to provide point
forecasts for a range of macroeconomic variables.  They are also
requested to provide probability forecasts, but these refer to annual,
rather than quarterly, percentage change, and thus are not considered
here.  The forecasts are issued in the middle of a quarter for the
current quarter (a prediction horizon of one quarter) and each of the
following four quarters (prediction horizons of two quarters through
five quarters).  While \cite{Engelberg&2009} show that the predictions
are mostly consistent with the hypothesis that SPF panel members
report their subjective means, medians or modes, they also note that
SPF forecasters tend to give more favorable views of the economy than
warranted by their subjective probabilities. 

Inflation rates based on the consumer price index (CPI) have been
included in the survey since the third quarter of 1981, and forecast
data are available on-line at \url{http://www.phil.frb.org/econ/spf/}.
As realizing observations, we use the CPI vintage from the Real-Time
Data Research Center of the Federal Reserve Bank of Philadelphia
\citep{CroushoreStark2001}, which is available from the third
quarter of 1994 on.  The traditional and probabilistic no-change
forecasts then use the most recent vintage available at
the issuing time of the forecast, while the predictive performance is
evaluated against the May 2010 vintage.  To convert the original
monthly observations of the CPI into annualized quarterly growth
rates, we follow common practice by averaging the monthly
observations of each quarter, and using the formula
\[ 
y_t = \Big( \Big( \hsp \frac{z_t}{z_{t-1}} \Big)^4 - 1 \Big) \times 100, 
\]
where $z_t$ is the observed quarterly CPI in quarter $t$, and $y_t$ is
the observed quarter-over-quarter growth rate of the CPI, or simply,
the inflation rate in quarter $t$, in percentage points.

We now compare the predictive performance of the SPF median forecast
to the traditional and probabilistic no-change forecasts.  Due to the
dynamic data vintage, the traditional no-change forecast does not
exactly trail the observations, even though it does so approximately.
The probabilistic no-change forecast uses the median of the 20 most
recent observations available at the issuing time.  Table
\ref{tab:MAE} summarizes the predictive performance for the period
from the third quarter of 1995 through the first quarter of 2010 in
terms of the mean absolute error (MAE).  In addition, we report the
results of \cite{DieboldMariano1995} tests of the hypothesis of equal
predictive performance between the SPF forecast and the 
reference forecasts.  In doing so, we provide the lower tail
probability under the null hypothesis in percentage points, where a
value of 00 indicates a lower tail probability less than or equal to
1\%, a value of 01 a lower tail probability between 1\% and 2\%,
\ldots, and a value of 99 a lower tail probability exceeding 99\%.
Thus, values from 00 to 04 and 95 to 99 correspond to a statistically
significant difference at the 5\% level for a one-sided test, and at
the 10\% level for a two-sided test.

The table confirms the well known fact that the SPF median forecast
outperforms the traditional no-change forecast at all prediction
horizons.  The corresponding MAE differentials are statistically
significant as well as substantial in size, reaching values close to a
percentage point.  At a nowcasting prediction horizon of a single
quarter, the SPF median forecast also has substantially lower MAE than
the probabilistic no-change forecast, even though the latter is more
competitive than the traditional no-change forecast.  Howeve, at prediction
horizons from two to five quarters, the SPF forecast is unable to
outperform the probabilistic no-change forecast, showing MAE values
that are about equal for the two methods.

\begin{table}[t]

\caption{Mean absolute error (MAE) for point forecasts of United
  States inflation from the third quarter of 1995 to the first quarter
  of 2010, in percentage points, along with the lower tail probability
  of the Diebold-Mariano test for equal predictive performance between
  the SPF forecast and the reference forecast.
  \label{tab:MAE}}

\begin{center} 
\begin{tabular}{ld{1.4}d{1.4}d{1.4}d{1.4}d{1.4}}
\toprule
Forecast method & \multicolumn{5}{c}{Prediction horizon in quarters} \\ 
& \multicolumn{1}{c}{1} & \multicolumn{1}{c}{2} 
& \multicolumn{1}{c}{3} & \multicolumn{1}{c}{4} & \multicolumn{1}{c}{5} \\
\midrule
SPF                     & 0.89     & 1.44      & 1.51     & 1.49     &  1.49 \\
Probabilistic no-change & 1.45_{99} & 1.46_{58} & 1.45_{18} & 1.48_{39} & 1.48_{44} \\
Traditional no-change   & 1.81_{99} & 2.06_{96} & 2.00_{91} & 2.06_{95} & 2.03_{97} \\
\bottomrule
\end{tabular} 
\end{center} 
\end{table}

Thus far, we have considered point forecasts, as opposed to
probabilistic forecasts or predictive distributions, which are of ever
increasing importance in a wide trans-disciplinary range of
applications \citep{Timmermann2000, Gneiting2008}.  The SPF and the
probabilistic no-change method provide predictive distributions in
natural ways, in that they can be identified with the discrete
probability measures that assign equal mass to each of the experts, or
each of the CPI observations in the training period, respectively.
For example, the predictive distribution that corresponds to our
standard version of the probabilistic no-change forecast assigns mass
$1/20$ to each of the 20 most recent inflation observations available
at the issuing time.  To obtain a predictive distribution associated
with the traditional no-change forecast, we take it to be Gaussian,
with mean equal to the most recent available observation, and variance
equal to the empirical mean squared error (MSE) of the traditional
no-change forecast over the rolling 20-quarter training period.

To assess the predictive performance of the probabilistic forecasts,
we use the continuous ranked probability score (CRPS), which is a
decision theoretically coherent proper scoring rule, and reduces to
the absolute error in the case of a point forecast
\citep{MathesonWinkler1976, GneitingRaftery2007}.  If the predictive
cumulative distribution function (CDF) is $F$ and the observation $y$
verifies, the CRPS is defined as
\begin{equation} \label{eq:CRPS} 
\textrm{CRPS}(F,y) 
= \int_{-\infty}^\infty \left( F(x) - \one\{x \geq y\} \right)^2 \: \dd x,
\end{equation} 
where $\one\{x \geq y\}$ denotes an indicator function that attains
the value $1$ if $x \geq y$ and the value $0$ otherwise.
\cite{Grimit&2006} showed that for a discrete probability measure $F$
that puts mass $1/M$ on each of $x_1, \ldots, x_M$, this can be
written as
\[
\textrm{CRPS}(F,y) = \frac{1}{M} \sum_{m=1}^M |x_m - y| - \frac{1}{M^2}
\sum_{m=1}^M \sum_{n=1}^M |x_m - x_n|. 
\]

\begin{table}[t]

\caption{Mean continuous ranked probability score (CRPS) for
  probabilistic forecasts of United States inflation from the third
  quarter of 1995 to the first quarter of 2010, in percentage points,
  along with the lower tail probability of the Diebold-Mariano test
  for equal predictive performance between the SPF forecast and the
  reference forecast. \label{tab:CRPS}}

\begin{center} 
\begin{tabular}{ld{1.4}d{1.4}d{1.4}d{1.4}d{1.4}}
\toprule
Forecast method & \multicolumn{5}{c}{Prediction horizon in quarters} \\ 
& \multicolumn{1}{c}{1} & \multicolumn{1}{c}{2} 
& \multicolumn{1}{c}{3} & \multicolumn{1}{c}{4} & \multicolumn{1}{c}{5} \\
\midrule
SPF                     & 0.69     &  1.16     & 1.25     & 1.26     & 1.27 \\
Probabilistic no-change & 1.08_{99} & 1.10_{21} & 1.10_{00} & 1.10_{00} & 1.11_{00} \\
Traditional no-change   & 1.56_{99} & 1.66_{97} & 1.51_{82} & 1.57_{90} & 1.47_{87} \\
\bottomrule
\end{tabular} 
\end{center} 

\end{table}

\begin{figure}[t]

\centerline{\includegraphics[width=8cm]{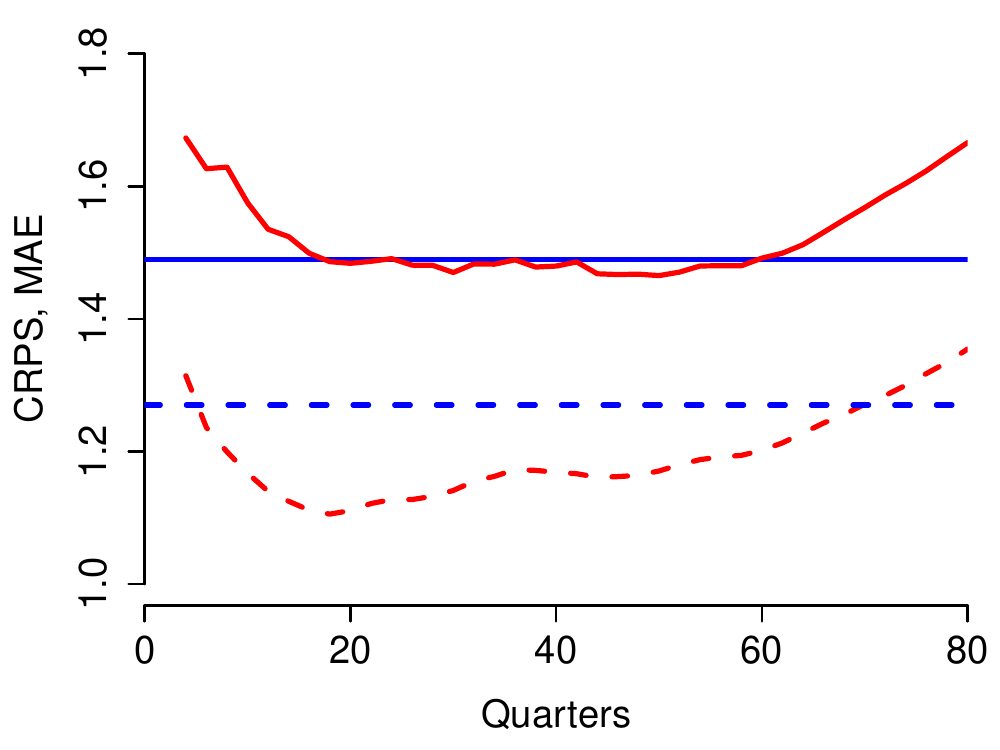}}

\vspace{-1mm}
\caption{Mean absolute error (MAE, red solid line) and mean
  continuous ranked probability score (CRPS, red dashed line) for
  five-quarter ahead probabilistic no-change forecasts for United
  States inflation from the third quarter of 1995 to the first quarter
  of 2010, as functions of the length of the rolling training period.
  For comparison, we also show the MAE (blue solid line) and the mean
  CRPS (blue dashed line) for the SPF forecast.
  \label{fig:PNC} }

\end{figure}

Table \ref{tab:CRPS} summarizes the predictive performance of the SPF,
probabilistic no-change and traditional no-change forecasts in terms
of the mean CRPS over the foregoing test period.  Again, the SPF
forecast outperforms the extant reference standard, namely the
traditional no-change forecast, at all prediction horizons.  In
nowcasts for the current quarter, the SPF forecast also shows
substantially lower CRPS than the probabilistic no-change forecast.
Beyond the current quarter in true forecasting mode the
simplistic probabilistic no-change forecast outperforms the SPF
experts.  Furthermore, as Figure \ref{fig:PNC} demonstrates, these
results are robust to the choice of the length of the rolling training
period for the probabilistic no-change forecast.

A valid concern at this point is that the SPF panel is simply a
collection of point forecasts that is not necessarily meant to be
taken as a discrete predictive distribution.  As noted, this resembles
the situation in weather forecasting where meteorologists use
ensembles of point forecasts from numerical weather prediction models
\citep{Palmer2002}, which are subject to biases and dispersion errors,
thus calling for statistical postprocessing
\citep{GneitingRaftery2005}.  Indeed, the Federal Reserve Bank of
Philadelphia has recently experimented with Gaussian density forecasts
that derive from the SPF panel \citep{Stark2010} and can be interpreted
as postprocessing methods.  In the next section, we take up this idea
and develop statistical postprocessing techniques that are tailored to
the SPF panel.

\section{Survey postprocessing}
\label{sec:postprocessing} 

In weather forecasting, statistical postprocessing methods have been
used with great success to improve the predictive performance of
ensemble prediction systems, with heteroscedastic regression
\citep{Gneiting&2005, ThorarinsdottirGneiting2010} and Bayesian model
averaging \citep{Raftery&2005, Sloughter&2010} being state of the art
techniques.  Here we introduce variants of these methods that are
tailored to the SPF ensemble and inflation forecasts.  In the SPF, the
composition of the expert panel changes gradually over time, with
individual members providing forecasts for about six years on average
\citep{Engelberg&2009}.  Furthermore, there are missing forecasts for
essentially all members even during the period when they are active.
Thus, it is very difficult to assess the predictive performance of
individual members, in contrast to what is commonly done in weather
forecasting.  Therefore, the predictive distributions obtained here depend
on summary statistics derived from the SPF panel or the probabilistic
no-change forecast.  In this context, we denote the median and the
variance of the SPF panel by $\mu_{\textrm{SPF}}$ and
$\sigma^2_{\textrm{SPF}}$.  Similarly, we write $\mu_{\textrm{PNC}}$
and $\sigma^2_{\textrm{PNC}}$ for the median and the variance
of the probabilistic no-change (PNC) forecast with a rolling 20
quarter training period.  The density function and the CDF of the
standard normal distribution will be denoted by $\varphi$ and $\Phi$,
respectively.

\subsection{Heteroscedastic regression}
\label{sec:HR} 

\cite{Gneiting&2005} and \cite{ThorarinsdottirGneiting2010} proposed a
statistical postprocessing method for ensembles of point forecasts
that uses heteroscedastic regression (HR) \citep{Leslie&2007}, where
the location parameter of the predictive distribution is a linear
function of the ensemble member forecasts, and the scale parameter is
a linear function of the ensemble variance.  Here, we adapt the method
so that the postprocessed predictive distribution is the 
asymmetric three-parameter two-piece normal distribution \citep{John1982} 
with density
\[
f_{\, \textup{TPN}}(y) = \begin{cases} 
\left( \displaystyle \frac{2}{\pi} \right)^{1/2} (\sigma_1+\sigma_2)^{-1} 
\exp \! \left( \displaystyle - \frac{(y-\mu)^2}{2\sigma_1^2} \right) & \textup{if} \quad y \leq \mu, \\
\left( \displaystyle \frac{2}{\pi} \right)^{1/2} (\sigma_1+\sigma_2)^{-1} 
\exp \! \left( \displaystyle - \frac{(y-\mu)^2}{2\sigma_2^2} \right) & \textup{if} \quad y \geq \mu, \rule{0mm}{9mm}
\end{cases}
\] 
and CDF
\begin{equation}  \label{eq:TPN} 
F_{\, \textup{TPN}}(y) = \begin{cases} 
\displaystyle 
\frac{2\sigma_1}{\sigma_1+\sigma_2} \, 
\Phi \Big( \frac{y-\mu}{\sigma_1} \Big) & \textup{if} \quad y \leq \mu, \\
\displaystyle 
\frac{\sigma_1-\sigma_2}{\sigma_1+\sigma_2} + \frac{2\sigma_2}{\sigma_1+\sigma_2} \,  
\Phi \Big( \frac{y-\mu}{\sigma_2} \Big) & \textup{if} \quad y \geq \mu. \rule{0mm}{9mm}
\end{cases}
\end{equation}
The two-piece normal distribution has been used for density forecasts
of United Kingdom inflation by the Bank of England since 1996
\citep{Wallis1999, Elder&2005}.  Its asymmetry allows for a distinct
treatment of upside and downside risks in inflation forecasting.  If
$\sigma_1 < \sigma_2$ the distribution is right skewed, and both the
mean and the median exceed its mode, $\mu$.  If $\sigma_1 > \sigma_2$
the distribution is left skewed.

We consider two variations of the HR approach.  The first variant models the
parameters of the two-piece normal distribution as functions of the
SPF median and SPF variance, in that
\begin{equation} \label{eq:HR1}
\mu = a + b \mu_{\textup{SPF}}, \quad 
\sigma_1^2 = c_1 + d_1 \sigma^2_{\textup{SPF}}, \quad 
\sigma_2^2 = c_2 + d_2 \sigma^2_{\textup{SPF}}, 
\end{equation}
and we refer to it as the HR model with SPF covariates.  The second
nests the first and posits
\begin{equation}\label{eq:HR2}
\mu = a + b_1 \mu_{\textup{SPF}} + b_2 \mu_{\textup{PNC}}, \quad 
\sigma_1^2 = c_1 + d_{11} \sigma^2_{\textup{SPF}} + d_{12} \sigma^2_{\textup{PNC}}, \quad
\sigma_2^2 = c_2 + d_{21} \sigma^2_{\textup{SPF}} + d_{22} \sigma^2_{\textup{PNC}}. 
\end{equation}
We refer to the specification in (\ref{eq:HR2}) as the HR model with
SPF and PNC covariates.  In out-of-sample forecasting overfitting is
heavily penalized, and it is an empirical question whether or not this
more complex model can outperform the more parsimonious one.

In \cite{Gneiting&2005} and \cite{ThorarinsdottirGneiting2010}, the
parameters of the postprocessing model were estimated by minimizing
the mean CRPS over the training period, and this was shown to yield
slightly better predictive performance than maximum likelihood
estimation.  In an economic context, similar approaches have been 
discussed by \cite{ElliottTimmermann2008}.  We adopt this
proposal, using formula (\ref{eq:CRPS.TPN}) in the Appendix, and
performing the minimization numerically, via the
Broyden-Fletcher-Goldfarb-Shanno algorithm as implemented in {\sc R} \citep{R2008}.
Each prediction horizon requires its individual fit with distinct
parameter estimates, using a rolling training period from the recent
past.

\begin{table}[t]

\caption{Minimum CRPS estimates for the HR density forecast
  (\ref{eq:HR2}) of United States inflation in the first quarter of
  2008 at a prediction horizon of two quarters, using a rolling
  training period of 40 quarters.  The resulting parameter values for
  the two-piece normal distribution (\ref{eq:TPN}) are mean value of 
  $\mu = 1.90$, with standard deviations
  $\sigma_1 = 0.59$ and $\sigma_2 = 3.27$, respectively.
  \label{tab:HR}}

\begin{center} 
\begin{tabular}{ccccccccc}
\toprule
$a$ & $b_1$ & $b_2$ & $c_1$ & $d_{11}$ & $d_{12}$ & $c_2$ & $d_{21}$ & $d_{22}$ \\ 
\midrule
0.36 & 0.53 & 0.00 & 0.00 & 0.52 & 0.00 & 0.00 & 0.00 & 3.10 \\
\bottomrule
\end{tabular} 
\end{center} 

\end{table}

\begin{figure}[t]

\centerline{\includegraphics[width=16cm]{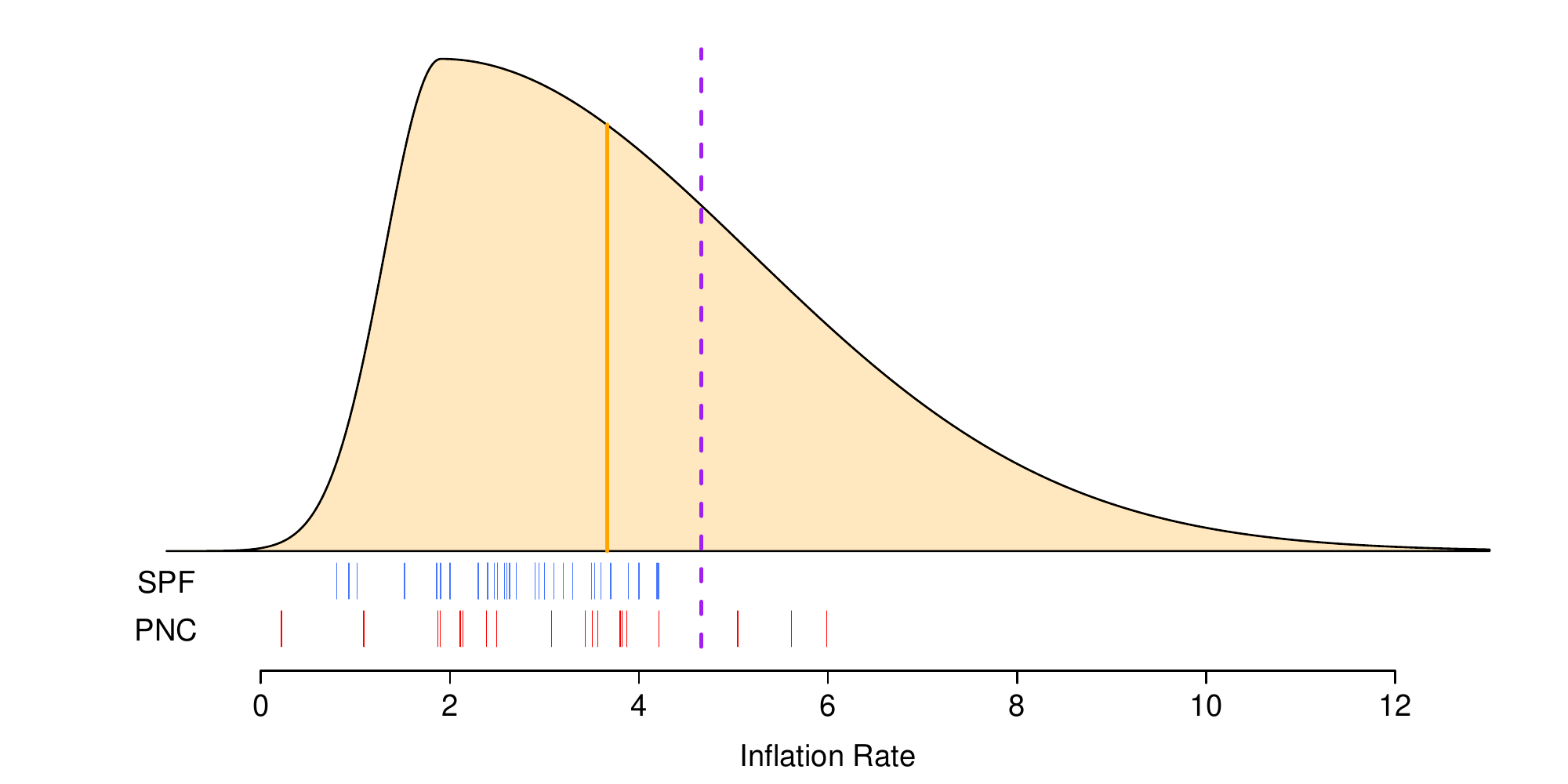}}

\vspace{-1mm}
\caption{Density forecast for the United States inflation rate in the
  first quarter of 2008 at a prediction horizon of two quarters,
  using the HR model (\ref{eq:HR2}) with both SPF and PNC covariates.
  At bottom, the SPF panel (blue bars), which has median 2.90, and the PNC
  ensemble (red bars), with median 3.25, are shown.  The median of the density
  forecast is indicated by the solid orange line at 3.67, and the realizing
  inflation rate is represented by the purple dashed line at 4.66.
  \label{fig:HR} }

\end{figure}

As an example, Figure \ref{fig:HR} and Table \ref{tab:HR} illustrate
the HR density forecast with SPF and PNC covariates for the first
quarter of 2008 at a prediction horizon of two quarters, fitted on a
rolling training period of 40 quarters.  The
point forecasts of the SPF experts had median 2.90 and standard
deviation 0.82.  The PNC ensemble had a higher median, at 3.30, and a
higher standard deviation, at 1.86.  Table \ref{tab:HR} shows the
parameter estimates for the HR model (\ref{eq:HR2}).  The mode $\mu$
of the two-piece normal distribution (\ref{eq:TPN}) is determined by
the SPF median, the downside risk $\sigma_1^2$ by the SPF spread, and
the upside risk $\sigma_2^2$ by the PNC spread.  Figure \ref{fig:HR}
shows the SPF and PNC ensembles along with the postprocessed density
forecast, which is strongly right skewed, with mode at 1.90 and median
at 3.67, higher than both the SPF median and the PNC median.  The
verifying inflation rate in the first quarter of 2008 was 4.66
percentage points.

\subsection{Gaussian mixture models}  \label{sec:GM}

Bayesian model averaging (BMA) is a standard method for combining
inferences \citep{Hoeting&1999}.  Its use in the statistical
postprocessing of ensemble weather forecasts was proposed by
\cite{Raftery&2005} and \cite{Sloughter&2010}.  In the normal mixture
version of \cite{Raftery&2005}, the BMA predictive density is a
mixture of Gaussian densities, where the components are associated
with individual ensemble member, and the mixture weights reflect the
members' relative contribution to predictive skill over the training
period.

Here we use a similar idea, taking the postprocessed predictive
distribution to be a mixture of two Gaussian components, with a CDF of
the form
\begin{equation}  \label{eq:GM}
F_{\, \textup{GM}}(y) = 
\alpha \, \Phi \! \left( \frac{y - \mu_1}{\sigma_1} \right) 
+ (1-\alpha) \, \Phi \! \left( \frac{y - \mu_2}{\sigma_2} \right) \! .  
\end{equation}
In a first variant we put 
\begin{equation} \label{eq:GM1}
\mu_1 = \mu_{\textup{SPF}}, \quad \sigma_1 = \sigma_{\textup{SPF}}, \quad 
\mu_2 = \mu_{\textup{PNC}}, \quad \sigma_2 = \sigma_{\textup{PNC}},   
\end{equation}
so that the mixture weight $\alpha \in [0,1]$ is the only parameter to
be estimated.  This is our most parsimonious, standard Gaussian
mixture (GM) model.  In a second variant, we put
\begin{equation} \label{eq:GM2}
\mu_1 = \mu_{\textup{SPF}}, \quad \mu_2 = \mu_{\textup{PNC}}, 
\end{equation}
with $\alpha \in [0,1]$, $\sigma_1 > 0$ and $\sigma_2 > 0$ to be
estimated.  We refer to (\ref{eq:GM2}) as the GM model with variance
adjustment.  In our experience, bias correction of the location
parameters deteriorates the predictive performance out of sample, 
and thus we do not
present results for such models.  The parameters are estimated by
maximum likelihood via the expectation-maximization (EM) algorithm,
based on training data from a rolling training period, as described by
\cite{Raftery&2005} and implemented in the {\sc R} package {\sc
  ensembleBMA}.

\begin{figure}[t]

\centerline{\includegraphics[width=16cm]{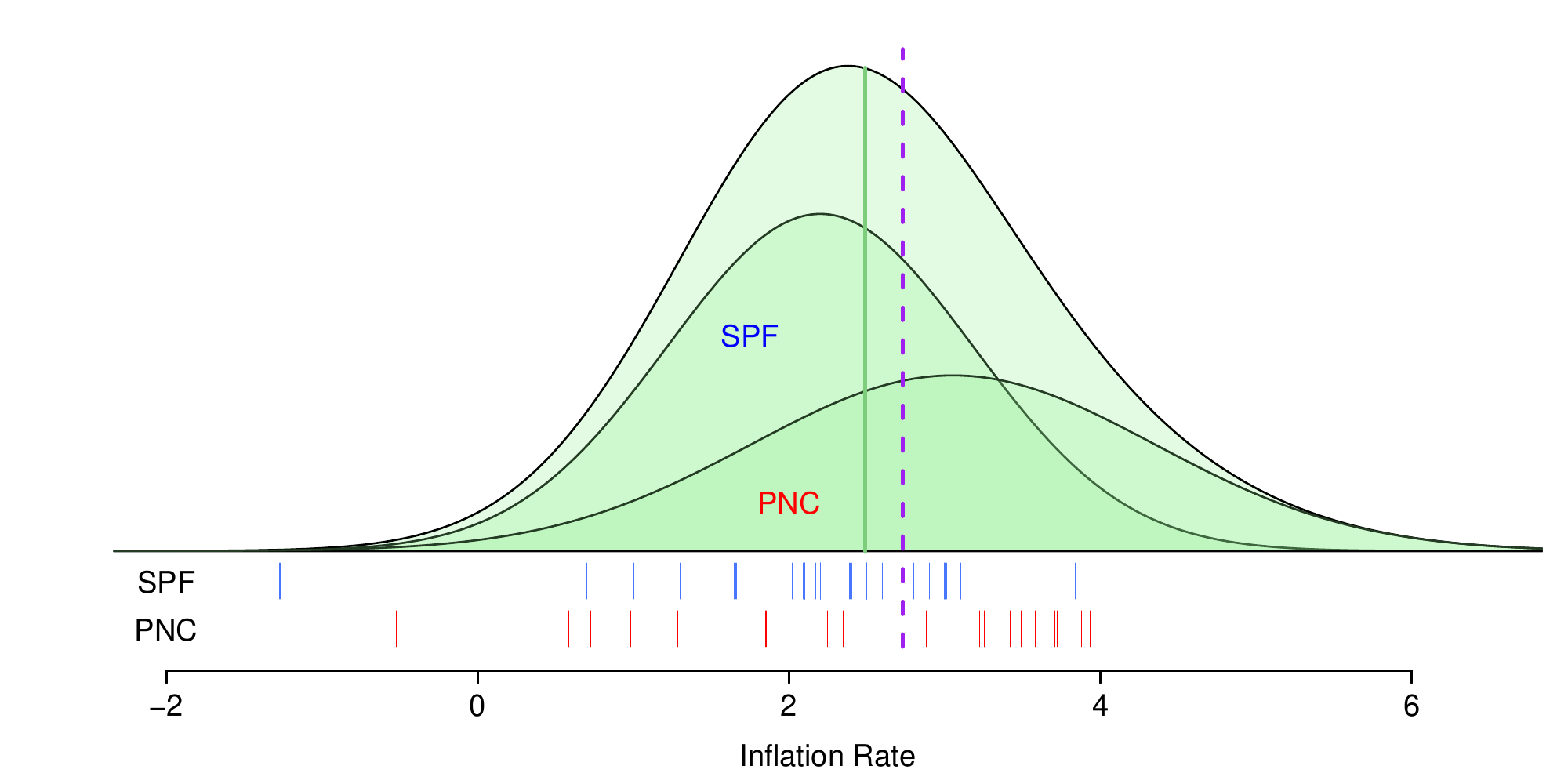}}

\vspace{-1mm}
\caption{Density forecast for the United States inflation rate in the
  second quarter of 2005 at a prediction horizon of two quarters (in pale green),
  using the GM model (\ref{eq:GM2}) with variance adjustment.  At
  bottom, the SPF panel (blue lines), which has median 2.20 and standard
  deviation 0.85, and the PNC ensemble (red lines), with median 3.05 and standard
  deviation 1.39, are shown.  The median of the density forecast is
  indicated by the green solid line at 2.49, and the realizing inflation
  rate is represented by the purple dashed line at 2.73.  The corresponding
  parameter estimates are $\alpha = 0.59$, $\sigma_1 = 0.98$ and
  $\sigma_2 = 1.30$.  The densities corresponding to the SPF panel and the PNC 
  forecast are indicated accordingly. 
  \label{fig:GM} }

\end{figure}

An example of a GM predictive distribution with variance adjustment is
given in Figure~\ref{fig:GM}, using a rolling training period of 40
quarters.

\subsection{Predictive performance}  \label{sec:results} 

In the SPF record, quarterly forecasts and observations of the CPI are
available for 115 quarters.  We split the data such that the forecasts
through the second quarter of 1995 are used solely for training
purposes, while the methods are tested on the data thereafter.  The 
CPI vintage data have been released quarterly since the third quarter of 
1994 and it would thus not have been possible to issue real-time forecasts 
based on this data prior to the third quarter of 1994.  For parameter estimation, we
use a rolling training period of 40 quarters.  The PNC ensembles used
are based on 20 quarterly observations.

\begin{figure}[t]

\centerline{\includegraphics[width=\textwidth]{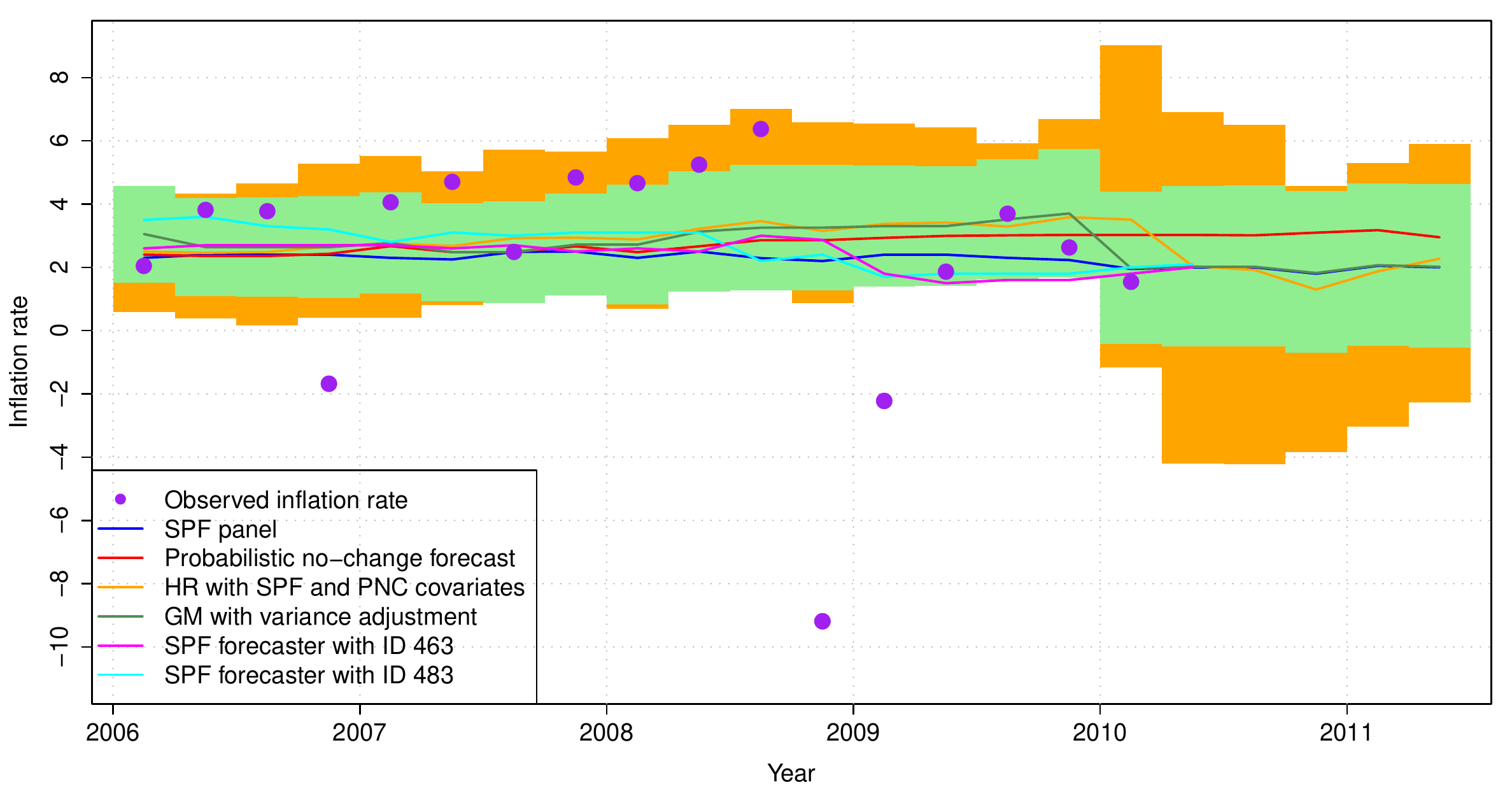}}

\vspace{-1mm}
\caption{Recent forecasts of the United States inflation rate at a
  prediction horizon of five quarters.  The point forecasts shown are
  the medians from the SPF panel data (blue line), the PNC ensemble
  (red line), the HR method with SPF and PNC covariates (orange
  line), and the GM technique with variance adjustment (green line).
  The 80\% prediction intervals for the HR method are indicated in orange,
  the 80\% prediction intervals for the GM technique are indicated in pale green,
  and the large purple dots represent the observed inflation rates.
  Furthermore, the point forecasts made by the individual SPF
  forecasters with ID 463 (magenta line) and ID 483 (cyan line) are
  shown.  Note that several of the forecasts are identical from early 2010 on. 
  \label{fig:forecasts} }

\end{figure}

Figure \ref{fig:forecasts} illustrates the postprocessed predictive
distributions using the HR method with SPF and PNC covariates, and the
GM technique with variance adjustment, for the period from 2006 on.
Through 2009, the methods yield very similar predictive distributions,
while their spread diverge for the period from early 2010 on.  This goes
hand-in-hand with the growing discrepancy between the median of the
SPF and the PNC ensembles.

In the performance comparison below, we include a probabilistic
forecast with conditional predictive CDF given by
\[
F(y) = 
\Phi \! \left( \frac{y - \mu_{\textup{SPF}}}{\textup{MSE}_{\textup{SPF}}} \right) \, 
\]
where $\textup{MSE}_{\textup{SPF}}$ denotes the mean squared error of
the SPF median over the past 40 quarters.  The Federal Reserve Bank of
Philadelphia has been experimenting with this method (Stark, 2010,
Section 3.2), \nocite{Stark2010} and we refer to it as the SPF median
with MSE technique.

Tables \ref{tab:MAEall} and \ref{tab:CRPSall} compare the predictive
performance of the postprocessing methods to that of the original SPF and
probabilistic no-change forecasts, in terms of the MAE and the mean
CRPS.  In nowcasting at a prediction horizon of one
quarter, none of the postprocessing techniques outperforms the SPF
forecast, neither in terms of the MAE nor the mean CRPS.  At
prediction horizons of two to five quarters, original and
postprocessed SPF forecasts as well as the probabilistic no-change
forecast all yield about equal MAE.  Postprocessing generally improves
the mean CRPS, reaching a predictive performance that is comparable,
but not superior, to that of the probabilistic no-change
forecast.

\begin{table}[p]

\caption{Mean absolute error (MAE) for forecasts of United States
  inflation from the third quarter of 1995 to the first quarter of
  2010, in percentage points, along with the lower tail probability of
  the Diebold-Mariano test under the hypothesis of the predictive
  performance being equal to that of the SPF forecast.  The 
  probabilistic no-change forecast (PNC) is obtained using 20 quarters
  of data, and the length of the training period for all
  postprocessing methods is 40 quarters.
  \label{tab:MAEall}}

\begin{center} 
\begin{tabular}{ld{1.4}d{1.4}d{1.4}d{1.4}d{1.4}}
\toprule
Forecast method & \multicolumn{5}{c}{Prediction horizon in quarters} \\ 
& \multicolumn{1}{c}{1} & \multicolumn{1}{c}{2} 
& \multicolumn{1}{c}{3} & \multicolumn{1}{c}{4} & \multicolumn{1}{c}{5} \\
\midrule
SPF                            & 0.89     & 1.44      & 1.51     & 1.49     & 1.49 \\
Probabilistic no-change        & 1.45_{99} & 1.46_{58} & 1.45_{18} & 1.48_{39} & 1.48_{44} \\ 
HR with SPF covariates         & 0.91_{60} & 1.42_{36} & 1.50_{44} & 1.53_{63} & 1.49_{50} \\
HR with SPF and PNC covariates & 0.87_{44} & 1.40_{27} & 1.51_{49} & 1.47_{40} & 1.48_{51} \\
GM                             & 1.05_{99} & 1.43_{47} & 1.45_{18} & 1.47_{36} & 1.47_{37} \\
GM with variance adjustment    & 0.89_{91} & 1.47_{69} & 1.50_{34} & 1.48_{39}  & 1.45_{16} \\
\bottomrule
\end{tabular} 
\end{center} 

\end{table}

\begin{table}[p]

\caption{Mean continuous ranked probability score (CRPS) for forecasts
  of United States inflation from the third quarter of 1995 to the
  first quarter of 2010, in percentage points, along with the lower
  tail probability of the Diebold-Mariano test under the hypothesis of
  the predictive performance being equal to that of the SPF forecast.
  The probabilistic no-change forecast (PNC) is obtained using 5 years
  of data and the length of the training period for all postprocessing
  methods is 40 quarters.
  \label{tab:CRPSall}}

\begin{center} 
\begin{tabular}{ld{1.4}d{1.4}d{1.4}d{1.4}d{1.4}}
\toprule
Forecast method & \multicolumn{5}{c}{Prediction horizon in quarters} \\ 
& \multicolumn{1}{c}{1} & \multicolumn{1}{c}{2} 
& \multicolumn{1}{c}{3} & \multicolumn{1}{c}{4} & \multicolumn{1}{c}{5} \\
\midrule
SPF                            & 0.69     & 1.16      & 1.25     & 1.26      & 1.27 \\
SPF median with MSE            & 0.70_{53} & 1.10_{39} & 1.15_{36} & 1.14_{34} & 1.13_{31} \\
Probabilistic no-change        & 1.08_{99} & 1.10_{21} & 1.10_{00} & 1.10_{00} & 1.11_{00} \\
HR with SPF covariates         & 0.69_{48} & 1.08_{09} & 1.15_{08} & 1.15_{12} & 1.13_{10} \\
HR with SPF and PNC covariates & 0.68_{40} & 1.09_{18} & 1.15_{12} & 1.14_{11} & 1.12_{12} \\
GM                             & 0.80_{96} & 1.09_{17} & 1.10_{01} & 1.11_{01} & 1.10_{00} \\
GM with variance adjustment    & 0.68_{36} & 1.10_{20} & 1.11_{00} & 1.12_{00} & 1.10_{00} \\
\bottomrule
\end{tabular} 
\end{center} 

\end{table}

\section{Robustness}  \label{sec:robust} 

We have argued that the SPF median forecast
outperforms simple no-change forecasts of United States inflation in
current quarter nowcasts.  However, at prediction horizons from two to
five quarters ahead, probabilistic no-change forecasts have equal or
higher skill than the SPF forecast, even after postprocessing.  In
making such claims, it is critically important to demonstrate the
robustness of the results under changes in the details of the
prediction experiment.

\begin{figure}[t]

\centerline{\includegraphics[width=8cm]{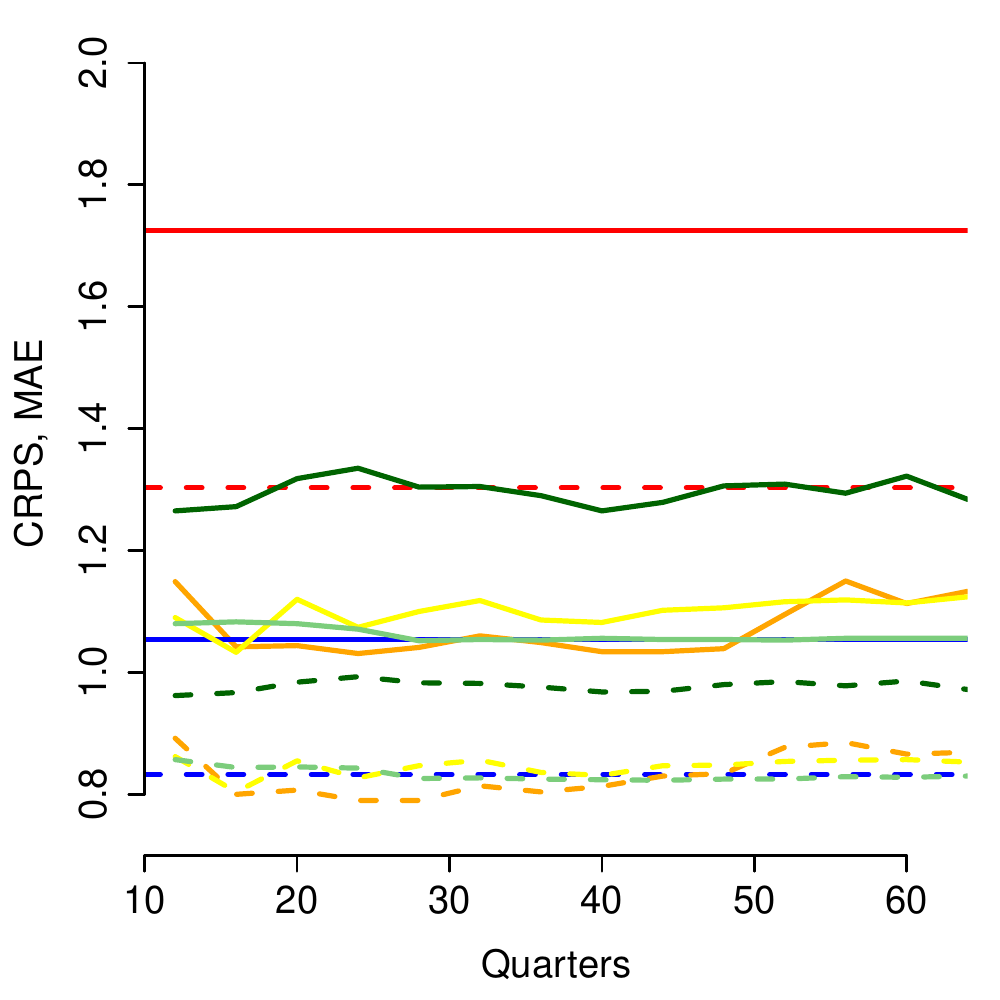}\includegraphics[width=8cm]{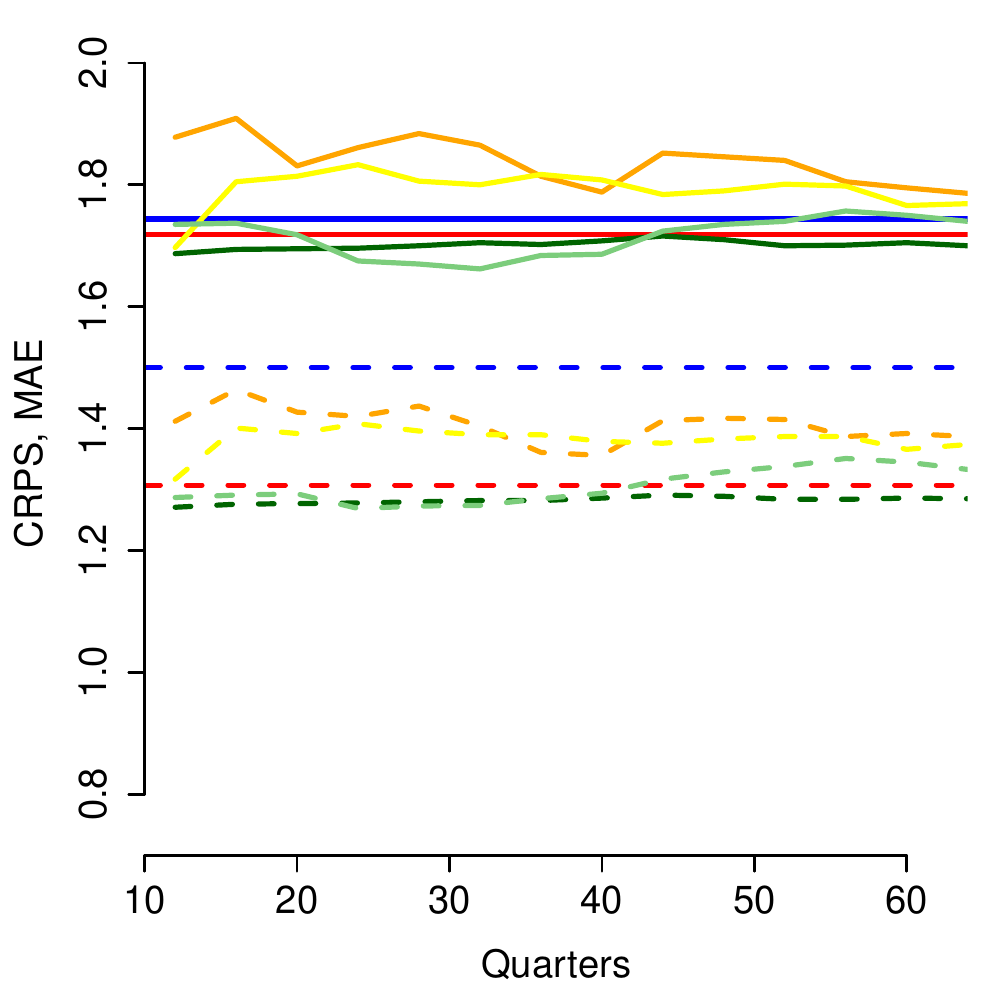}}

\vspace{-1mm}
\caption{MAE (solid lines) and mean CRPS (dashed lines) for forecasts
  of United States inflation from the first quarter of 2000 to the
  first quarter of 2010, as a function of the length of the rolling
  training period, at prediction horizons of one quarter (left) and
  five quarters (right).  Results are shown for the SPF (blue),
  probabilistic no-change (red), HR with SPF covariates (yellow), HR
  with SPF and PNC covariates (orange), GM (dark green), and GM with
  variance adjustment (pale green) techniques.
   \label{fig:training} }

\end{figure}

An initial robustness check was done in Section \ref{sec:reference},
where Figure \ref{fig:PNC} showed the performance of the probabilistic
no-change forecast to depend little on the choice of the length of the
rolling training period, thereby justifying that we fix it at 20
quarters.  In this section we show that our key findings remain valid
under changes in the length of the training period for the
postprocessing techniques, and we check whether they hold in smaller
test periods, under distinct economic regimes.

Figure \ref{fig:training} plots the MAE and the mean CRPS for our
postprocessing methods in their dependence on the length of the
rolling training period, at prediction horizons of one and five
quarters ahead.  Except for the probabilistic no-change forecast, all
methods show substantially higher skill at the shorter prediction
horizon.  Furthermore, for all methods the mean CRPS is much lower
than the MAE.  The predictive skill of the postprocessing methods does
not depend much on the length of the training period, and our choice
of a 40 quarter period seems reasonable.

Next we assess the effect of the test period and the corresponding
economic regimes.  In Tables \ref{tab:MAE:all} and \ref{tab:CRPS:all},
the aggregate results in Tables \ref{tab:MAE} and \ref{tab:CRPS},
which cover the third quarter of 1995 through the first quarter of
2010, have been stratified into three sub-periods of about equal
length.  With shorter test periods, the Diebold-Mariano test statistic
is occasionally ill defined because of a negative variance estimate.
\cite{DieboldMariano1995} suggest that the variance estimate should
then be treated as zero and the null hypothesis of equal forecast
accuracy be rejected.

The first sub-period ranges from the third quarter of 1995 through the
fourth quarter of 2000.  This was an era of general economic boom in
the United States and inflation rates were particularly stable, which
facilitated forecasting and is mirrored in low MAEs.  Hence, the
traditional no-change forecast performed quite well during this
period, particularly at prediction horizons of three and four
quarters, where it had lower MAE than all other forecasts, including
the SPF and the probabilistic no-change forecasts.  Another look at
Figure \ref{fig:SPF} supports this choice, because the behavior of
inflation rates in the late 1990s appears generally compatible with a
symmetric random walk model, and the traditional no-change forecast is
the Bayes predictor under such an assumption \citep{Granger1969,
  Gneiting2010b}.  Thus, during this period of sustained economic
growth the choice of the traditional no-change forecast as reference
standard was appropriate.

\begin{table}[p]

\caption{Mean absolute error (MAE) for forecasts of United States
  inflation from the third quarter of 1995 to the fourth quarter of
  2000 (top panel), from the first quarter of 2001 to the fourth
  quarter of 2005 (middle panel), and from the first quarter of 2006
  to the first quarter of 2010 (bottom panel), in percentage points,
  along with the lower tail probability of the Diebold-Mariano test
  under the hypothesis of the predictive performance being equal to
  that of the SPF forecast.  The NA symbol indicates a missing value
  due to a negative variance estimate.
  \label{tab:MAE:all}}

\begin{center} 
\begin{tabular}{ld{1.4}d{1.4}d{1.4}d{1.4}d{1.4}}
\toprule
Forecast method & \multicolumn{5}{c}{Prediction horizon in quarters} \\ 
& \multicolumn{1}{c}{1} & \multicolumn{1}{c}{2} 
& \multicolumn{1}{c}{3} & \multicolumn{1}{c}{4} & \multicolumn{1}{c}{5} \\
\midrule
1995:Q3 - 2000:Q4 & & & & & \\
\midrule
SPF                            & 0.52      & 0.87     & 0.93     & 0.95      & 0.95 \\
Probabilistic no-change        & 0.87_{99} & 0.90_{72} & 0.93_{49} & 0.96\na  & 0.98\na \\
Traditional no-change          & 0.76_{99} & 0.92_{66} & 0.78_{14} & 0.88_{33} & 1.15_{94} \\
HR with SPF covariates         & 0.54_{65} & 0.84_{37} & 0.94_{52} & 1.00_{57} & 0.99_{55} \\
HR with SPF and PNC covariates & 0.56_{75} & 0.88_{56} & 0.96_{60} & 1.03_{62} & 1.01_{58} \\
GM                             & 0.59_{99} & 0.91_{90} & 0.94_{67} & 0.96\na  & 0.97\na \\
GM with variance adjustment    & 0.52_{99} & 0.87_{92} & 0.93_{17} & 0.95\na  & 0.95_{91} \\
\midrule 
2001:Q1 - 2005:Q4 & & & & & \\
\midrule
SPF                            & 1.00      & 1.26     & 1.33     & 1.26      & 1.28 \\
Probabilistic no-change        & 1.18_{87} & 1.18_{12} & 1.17_{10} & 1.16_{24} & 1.13_{01} \\
Traditional no-change          & 1.65_{99} & 1.52_{90} & 1.74_{97} & 1.86_{99} & 1.58_{99} \\
HR with SPF covariates         & 1.11_{98} & 1.20_{03} & 1.25_{22} & 1.22_{36} & 1.18_{00} \\
HR with SPF and PNC covariates & 1.13_{99} & 1.21_{11} & 1.25_{25} & 1.22_{37} & 1.17_{00} \\
GM                             & 1.08_{83} & 1.17_{10} & 1.16_{08} & 1.15_{20} & 1.14_{01} \\
GM with variance adjustment    & 1.00_{83} & 1.24_{00} & 1.29_{09} & 1.19_{13} & 1.17_{11} \\
\midrule
2006:Q1 - 2010:Q1 & & & & & \\
\midrule
SPF                            & 1.22      & 2.39     & 2.48     & 2.47      & 2.45 \\
Probabilistic no-change        & 2.52_{99} & 2.51_{65} & 2.46_{45} & 2.52_{70} & 2.55_{69} \\
Traditional no-change          & 3.36_{99} & 4.18_{96} & 3.88_{90} & 3.84_{92} & 3.72_{94} \\
HR with SPF covariates         & 1.14_{38} & 2.43_{60} & 2.53_{66} & 2.59_{78} & 2.52_{77} \\
HR with SPF and PNC covariates & 1.00_{20} & 2.33_{38} & 2.51_{57} & 2.35_{00} & 2.51_{62} \\
GM                             & 1.62_{97} & 2.43_{56} & 2.46_{45} & 2.52_{69} & 2.52_{65} \\
GM with variance adjustment    & 1.22_{84} & 2.52_{73} & 2.48_{50} & 2.52_{71}  & 2.44_{48} \\
\midrule
SPF forecaster ID 463          & 1.06_{16} & 2.35_{34} & 2.57_{75} & 2.41_{13} & 2.33_{00} \\
SPF forecaster ID 483          & 2.12_{99} & 2.25_{22} & 2.30_{15} & 2.26_{05} & 2.33_{19} \\
SPF forecaster ID 510          & 2.32_{99} & 2.67_{98} & 2.48_{50} & 2.71\na  & 2.55_{96} \\
SPF forecaster ID 516          & 1.53_{83} & 3.34_{99} & 3.03\na  & 2.78_{90} & 2.94_{99} \\
SPF forecaster ID 528          & 1.40_{68} & 2.59_{76} & 3.00_{98} & 2.70\na  & 2.55_{90} \\
\bottomrule
\end{tabular} 
\end{center} 

\end{table}

The second sub-period extends from the first quarter of 2001 through
the fourth quarter of 2005.  The expansion that began in the early
1990s came to an end at the beginning of this period with the early
2000s recession, and inflation rates started to behave much more
erratically, not unlike a white noise model.  The erratic behavior
becomes even more pronounced during the third sub-period, which ranges from
the first quarter of 2006 through the first quarter of 2010.  This
includes the late 2000s recession, a financial crisis much more severe
than that of the previous period.  Its effect culminated in the
extreme deflation observed in the fourth quarter of 2008, at $- 9.2$
percentage points, and we now see substantially lower levels of
predictability than before.

Generally, from 2000 on the probabilistic no-change forecast succeeds
the traditional no-change forecast as the appropriate reference
forecast.  In line with the aggregate results in Tables \ref{tab:MAE}
and \ref{tab:CRPS}, at prediction horizons from two to five quarters
the SPF forecast tends to benefit from postprocessing in terms of the
mean CRPS, but not in terms of the MAE.  We conclude that the spread
adjustment generally is useful, but not the location adjustment.
However, while the probabilistic forecast performance improves under
postprocessing as measured by the mean CRPS, the postprocessed forecasts 
are unable to outperform the simplistic probabilistic no-change forecast.

The sub-period from 2006 on also is the longest consecutive period for
which a complete set of forecasts is available for more than one
individual SPF expert.  In Table \ref{tab:MAE} we show MAEs for the
five professionals who had complete records during this period.  The
predictive performance of the experts with IDs 463 and 483 is
particularly impressive, and their point forecasts are illustrated in
Figure \ref{fig:forecasts}.

\begin{table}[p]

\caption{Mean continuous ranked probability score (CRPS) for forecasts
  of United States inflation from the third quarter of 1995 to the
  fourth quarter of 2000 (top panel), from the first quarter of 2001
  to the fourth quarter of 2005 (middle panel), and from the first
  quarter of 2006 to the first quarter of 2010 (bottom panel), in
  percentage points, along with the lower tail probability of the
  Diebold-Mariano test under the null hypothesis of equal predictive
  performance between the SPF forecast and the reference forecast.
  The NA symbol indicates a missing value due to a negative variance
  estimate.
  \label{tab:CRPS:all}}

\begin{center} 
\begin{tabular}{ld{1.4}d{1.4}d{1.4}d{1.4}d{1.4}}
\toprule
Forecast method & \multicolumn{5}{c}{Prediction horizon in quarters} \\ 
& \multicolumn{1}{c}{1} & \multicolumn{1}{c}{2} 
& \multicolumn{1}{c}{3} & \multicolumn{1}{c}{4} & \multicolumn{1}{c}{5} \\
\midrule
1995:Q3 - 2000:Q4  & & & & \\ 
\midrule
SPF                            & 0.39     & 0.66      & 0.73     & 0.76     & 0.77 \\
SPF median with MSE            & 0.39_{38} & 0.59_{04} & 0.65_{12} & 0.71_{24} & 0.72_{30} \\
Probabilistic no-change        & 0.59_{99} & 0.62_{20} & 0.63_{01} & 0.66_{00} & 0.68_{01} \\
Traditional no-change with MSE & 0.55_{98} & 0.63_{36} & 0.57_{10} & 0.70_{31} & 0.86_{81} \\
HR with SPF covariates         & 0.40_{62} & 0.60_{21} & 0.66_{26} & 0.73_{44} & 0.74_{44} \\
HR with SPF and PNC covariates & 0.42_{76} & 0.63_{34} & 0.68_{30} & 0.75_{48} & 0.76_{48} \\
GM                             & 0.43_{98} & 0.63_{20} & 0.66_{01} & 0.67_{00} & 0.69_{00} \\
GM with variance adjustment    & 0.38_{22} & 0.58_{00} & 0.61_{00} & 0.64_{00} & 0.66_{02} \\
\midrule 
2001:Q1 - 2005:Q4 & & & & \\
\midrule
SPF                            & 0.82     & 1.03      & 1.07     & 1.07     & 1.07 \\
SPF median with MSE            & 0.82_{59} & 0.94_{01} & 0.97_{00} & 0.93_{00} & 0.92_{00} \\
Probabilistic no-change        & 0.87_{73} & 0.88_{01} & 0.89_{00} & 0.89_{00} & 0.87_{00} \\
Traditional no-change with MSE & 1.17_{98} & 1.20_{86} & 1.26_{83} & 1.41_{96} & 1.15_{72} \\
HR with SPF covariates         & 0.85_{78} & 0.93_{02} & 0.97_{07} & 0.94_{06} & 0.88_{00} \\
HR with SPF and PNC covariates & 0.88_{90} & 0.94_{03} & 0.98_{09} & 0.95_{06} & 0.87_{00} \\
GM                             & 0.81_{44} & 0.88_{01} & 0.86_{00} & 0.86_{02} & 0.83_{00} \\
GM with variance adjustment    & 0.80_{21} & 0.92_{00} & 0.94_{00} & 0.91_{00} & 0.85_{00} \\
\midrule
2006:Q1 - 2010:Q1 & & & & \\
\midrule
SPF                            & 0.92     & 1.98      & 2.13     & 2.15     & 2.17 \\
SPF median with MSE            & 0.95_{71} & 1.96_{46} & 2.02_{32} & 1.96\na  & 1.90\na \\ 
Probabilistic no-change        & 1.95_{99} & 1.98_{50} & 1.94_{09} & 1.95_{08} & 1.95\na \\
Traditional no-change with MSE & 3.31_{99} & 3.54_{98} & 3.03_{84} & 2.87_{85} & 2.64_{79} \\
HR with SPF covariates         & 0.86_{39} & 1.87_{29} & 1.99_{22} & 1.95_{11} & 1.95\na \\
HR with SPF and PNC covariates & 0.78_{21} & 1.86_{31} & 1.97_{24} & 1.91_{13} & 1.91\na \\
GM                             & 1.26_{97} & 1.93_{42} & 1.96_{15} & 1.96_{15} & 1.95_{12} \\
GM with variance adjustment    & 0.93_{58} & 1.99_{52} & 1.96_{13} & 1.99_{12} & 1.95_{07} \\
\bottomrule
\end{tabular} 
\end{center} 

\end{table}

\section{Conclusions}  \label{sec:discussion} 

Predicting inflation is important, and there are various ways of doing
it, including forecasts from the Survey of Professional Forecasters
(SPF) and simple no-change forecasts.  In the extant literature, the
traditional no-change forecast has served as a benchmark, to which
more sophisticated techniques are to be compared.  While this was
appropriate during the late 1990s economic boom, in today's turbulent
markets the equally simplistic probabilistic no-change forecast
performs much better.  To avoid spurious claims of predictability, we
make a plea for the use of the probabilistic no-change forecast as a
default reference standard in inflation forecasting.

In current quarter nowcasting, corresponding to a prediction horizon
of one quarter, economic expertise is persuasive, and the SPF professionals
outperform all types of no-change forecasts.  Businesses,
organizations and the government are well advised to avail themselves
of the SPF experts' short-term predictions.

At prediction horizons beyond the current quarter, the probabilistic
predictive performance of the SPF forecast, as measured by the mean
CRPS, improves under statistical postprocessing.  Supplementing the
SPF median with MSE method, which has been used by the Federal Reserve
Bank of Philadelphia, we have introduced
heteroscedastic regression (HR) and Gaussian mixture (GM) techniques
for doing this.  However, at prediction horizons of two or more
quarters, even postprocessed SPF forecasts fail to outperform the
simplistic probabilistic no-change forecast.

While novel and potentially surprising in the specific context of the
SPF and inflation rates, the result conforms with a general theme in
the forecasting literature, in that simple prediction methods tend to
perform well, with overfitting being heavily penalized and subjective
human expertise often being overrated.  For example, \cite{Nelson1972}
showed that simple statistical methods can outperform complex economic
simulation models, and a recent 20-year study argues persuasively that
professional forecasters, who appear as experts in the media and
advise governments, organizations and businesses, might not be better
prognosticians than John Q.~Public \citep{Tetlock2005}.  In this light, 
we might be well advised to adapt to low levels of socio-economic
predictability, where uncertainty reigns \citep{MakridakisTaleb2009}.

\section*{Appendix}

Here we provide a closed form expression for the continuous ranked
probability score (\ref{eq:CRPS}) when the predictive distribution is
the three-parameter two-piece normal distribution with CDF given by
(\ref{eq:TPN}).  A tedious but straightforward calculation shows that
\begin{equation}  \label{eq:CRPS.TPN}
\textrm{CRPS}(F_{\textrm{TPN}}, y) = \begin{dcases}
\frac{4\sigma_1^2}{\sigma_1+\sigma_2} \; \Big[ \frac{y-\mu}{\sigma_1} \,
\Phi \Big( \frac{y-\mu}{\sigma_1} \Big) + \varphi \Big( \frac{y-\mu}{\sigma_1} \Big) \Big] & \\
\qquad - \; (y-\mu) & \\
\qquad + \; \frac{2}{\sqrt{\pi}} \, \frac{\sqrt{2} \sigma_2 (\sigma_2^2-\sigma_1^2) - (\sigma_1^3+\sigma_2^3)}
{(\sigma_1+\sigma_2)^2} & \textup{if} \quad y \leq \mu, \\
\frac{4\sigma_2^2}{\sigma_1+\sigma_2} \; \Big[ \frac{y-\mu}{\sigma_2} \,
\Phi \Big( \frac{y-\mu}{\sigma_2} \Big) + \varphi \Big( \frac{y-\mu}{\sigma_2} \Big) \Big] & \rule{0mm}{9mm} \\
\qquad + \; \frac{(\sigma_1-\sigma_2)^2-4\sigma_2^2}{(\sigma_1+\sigma_2)^2} \, (y-\mu) & \\
\qquad + \; \frac{2}{\sqrt{\pi}} \, \frac{\sqrt{2} \sigma_1 (\sigma_1^2-\sigma_2^2) - (\sigma_1^3+\sigma_2^3)}
{(\sigma_1+\sigma_2)^2} & \textup{if} \quad y \geq \mu.
\end{dcases}
\end{equation}
\cite{Grimit&2006} give a similar formula when the predictive
distribution is a mixture of Gaussian components, which we use to
compute the CRPS for the GM forecasts.

\section*{Acknowledgments}

We are grateful to Tom Stark for generously sharing his knowledge
about the SPF data, and to Marc Genton and Alex Lenkoski for discussions.  
Funding was provided by NSF grant DMS-0706745 to the University of 
Washington and by the Alfried Krupp von Bohlen und Halbach-Foundation.

\bibstyle{jasa}
\bibliography{econ}

\end{document}